\documentclass[pra,twocolumn,superscriptaddress]{revtex4-1}

\usepackage[utf8]{inputenc}
\usepackage[english]{babel}
\usepackage[T1]{fontenc}
\usepackage{lmodern}

\usepackage{graphicx}
\usepackage{amsmath}
\usepackage{amsfonts}
\usepackage{amssymb}
\usepackage{amsthm}
\usepackage{color}

\usepackage{mathtools}

\usepackage{siunitx}

\usepackage{dsfont}
\usepackage{microtype}

\usepackage[breaklinks=true,colorlinks]{hyperref}

\newcommand{\cf}{cf.\ }

\newcommand{\bd}{\begin{equation*}}
\newcommand{\ed}{\end{equation*}}

\begin{document}

\title{Two-parameter counter-diabatic driving in quantum annealing}

\author{Luise Prielinger}
\affiliation{Atominstitut, TU Wien, Stadionallee 2, 1020 Vienna, Austria}
\author{Andreas Hartmann}
\email{Corresponding author. andreas.hartmann@uibk.ac.at}
\affiliation{Institut f\"ur Theoretische Physik, Universit\"at Innsbruck, Technikerstra{\ss}e~21a, A-6020~Innsbruck, Austria}
\author{Yu Yamashiro}
\affiliation{Department of Physics, Tokyo Institute of Technology, Nagatsuta-cho, Midori-ku, Yokohama 226-8503, Japan}
\affiliation{Jij Inc., Bunkyo-ku, Tokyo 113-0031, Japan}
\author{Kohji Nishimura}
\affiliation{Jij Inc., Bunkyo-ku, Tokyo 113-0031, Japan}
\author{Wolfgang Lechner}
\affiliation{Institut f\"ur Theoretische Physik, Universit\"at Innsbruck, Technikerstra{\ss}e~21a, A-6020~Innsbruck, Austria}
\affiliation{Parity Quantum Computing GmbH, Rennweg 1, A-6020~Innsbruck, Austria}
\author{Hidetoshi Nishimori}
\affiliation{Institute of Innovative Research, Tokyo Institute of Technology, Nagatsuta-cho, Midori-ku, Yokohama 226-8503, Japan}
\affiliation{Graduate School of Information Sciences, Tohoku University, Sendai 980-8579, Japan}
\affiliation{RIKEN Interdisciplinary Theoretical and Mathematical Sciences Program (iTHEMS), Wako, Saitama 351-0198, Japan}

\begin{abstract}
We introduce a two-parameter approximate counter-diabatic term into the Hamiltonian of the transverse-field Ising model for quantum annealing to accelerate convergence to the solution, generalizing an existing single-parameter approach. The protocol is equivalent to unconventional diabatic control of the longitudinal and transverse fields in the transverse-field Ising model and thus makes it more feasible for experimental realization than an introduction of new terms such as non-stoquastic catalysts toward the same goal of performance enhancement. We test the idea for the $p$-spin model with $p=3$, which has a first-order quantum phase transition, and show that our two-parameter approach leads to significantly larger ground-state fidelity and lower residual energy than those by traditional quantum annealing and by the single-parameter method.  We also find a scaling advantage in terms of the time-to-solution as a function of the system size in a certain range of parameters as compared to the traditional methods in the sense that an exponential time complexity is reduced to another exponential complexity with a smaller coefficient. Although the present method may not always lead to a drastic exponential speedup in difficult optimization problems, it is useful because of its versatility and applicability for any problem after a simple algebraic manipulation, in contrast to some other powerful prescriptions for acceleration such as non-stoquastic catalysts in which one should carefully study in advance if it works in a given problem and should identify a proper way to meticulously control the system parameters to achieve the goal, which is generally highly non-trivial.

\end{abstract}

\date{\today}

\maketitle

\section{Introduction}
Quantum annealing is a metaheuristic for combinatorial optimization problems~\cite{kadowaki1998quantum, brooke1999quantum, santoro2002theory, santoro2006optimization, das2008colloquium, morita2008mathematical, hauke2020perspectives} and has often been analyzed theoretically in the framework of adiabatic quantum computing~\cite{farhi2000quantum, farhi2001quantum, albash2018adiabatic}. A serious bottleneck in this approach originates in the exponential closing of the energy gap between the ground state and the first excited state as a function of the system size, typically at a first-order quantum phase transition, by which the computation time explodes exponentially according to the adiabatic theorem of quantum mechanics (see, e.g., Ref.~\cite{albash2018adiabatic}). 
One of the promising candidates to circumvent this difficulty is diabatic quantum annealing~\cite{crosson2020prospects}, in which one ingeniously drives the system out of the ground state to avoid the problem of a closing minimal energy gap and thus to reach the final ground state with high fidelity. There have been attempts to design protocols to control the system variables based on this idea~\cite{crosson2020prospects}, and shortcuts to adiabaticity~\cite{arimondo2013chapter, delcampo2015controlling, delcampo2019focus, guery-odelin2019shortcuts} present strong candidates, providing a systematic way toward this goal.

\par 

Among these shortcuts-to-adiabaticity methods~\cite{demirplak2003adiabatic, berry2009transitionless, chen2010fast, chen2011lewisriesenfeld, takahashi2013transitionless, jarzynski2013generating, takhashi2019hamiltonian}, counter-diabatic (CD) driving~\cite{delcampo2013shortcuts, jarzynski2013generating, damski2014counterdiabatic, sels2017minimizing, claeys2019floquet, hartmann2019rapid} is one of the most promising approaches. The underlying idea of CD driving is to speed up an originally-adiabatic process by additionally applying a CD Hamiltonian (adiabatic gauge potential) that suppresses the transitions between the system eigenstates. However, for many-body quantum systems, finding the exact CD Hamiltonian requires {\em a priori} knowledge of these eigenstates at all times during the dynamics~\cite{berry2009transitionless}, which is practically unfeasible. Recently, Sels, Polkovnikov, and collaborators~\cite{sels2017minimizing, kolodrubetz2017geometry, claeys2019floquet} have developed a variational approach where a simple and local, but approximate, CD Hamiltonian is introduced, which makes the formulation and realization much simpler not just theoretically but experimentally as well~\cite{zhou2020experimental, hartmann2019rapid} (see, also,~\cite{passarelli2020counterdiabatic, hegade2021shortcuts} for related developments). The price to pay is that the enhancement of performance is often limited.

\par 

In the present contribution, we propose a method to identify an enhanced local approximate CD Hamiltonian. The latter entails a second adiabatic gauge potential that appears naturally due to the introduction of an additional time-dependent driving function of the Hamiltonian. We find its optimal coefficients by minimizing the operator distance between the exact and approximate CD Hamiltonians in order to maximize the performance of the latter. This approach generalizes the existing method of single-parameter local CD driving by expanding the search space into a second dimension.
We test the idea for the $p$-spin model with $p=3$ as the problem Hamiltonian, which is known to be a simple model, yet a hard problem to solve by traditional quantum annealing~\cite{joerg2010energy, seki2012quantum, seoane2012manybody, seki2015quantum, nishimori2017exponential, hartmann2019quantum}.
We demonstrate that our approximate two-parameter CD Hamiltonian leads to clearly enhanced final ground-state fidelity and reduced residual energy compared to traditional quantum annealing and the existing method of the approximate single-parameter CD Hamiltonian. We further show a scaling advantage of the method  compared to its traditional counterparts in a certain parameter range in the sense that an exponential time complexity is reduced to another exponential complexity with a smaller coefficient. Our two-parameter CD Hamiltonian improves the ground-state fidelity and residual energy for both short and longer annealing times which thus decreases the time-to-solution considerably. We note that the modified CD Hamiltonian used in this approach involves only local $\sigma^y_i$ operators, where $i$ is the site index, and can thus be rotated in the spin space at each site such that the result consists only of $\sigma_i^x$ and $\sigma_i^z$ in addition to the original transverse-field Ising Hamiltonian. This is simply the usual transverse-field Ising model, but with unconventional diabatic control of the transverse and longitudinal fields, making it feasible for experimental realization.

\par 

The paper is structured as follows. In Sec.~\ref{sec_methods}, we introduce the method of finding the two-parameter CD protocol and apply the formulation to the $p$-spin model. Numerical tests are presented in Sec.~\ref{sec_results} for the $p$-spin model with $p=3$, and Sec.~\ref{sec_discussion} discusses and concludes the paper.

\section{Method}\label{sec_methods}
Quantum annealing is a metaheuristic that aims to solve combinatorial optimization problems. The basic idea is to find the lowest-energy eigenstate of a problem Hamiltonian $\mathcal{H}_\mathrm{p}$--- that encodes a combinatorial optimization problem that we want to solve as an Ising model~\cite{lucas2014ising}--- by adiabatically transferring the easy-to-prepare ground state of the driver Hamiltonian
\begin{align}
    \mathcal{H}_\mathrm{d}=-\gamma\sum_{i=1}^N \sigma_i^x
    \label{eq_H_driver}
\end{align}
with $\gamma$ the time-independent transverse magnetic field strength and $N$ the number of sites (qubits) in the system, into the ground state of $\mathcal{H}_\mathrm{p}$.
The annealing schedule is often chosen as
\begin{equation}
\mathcal{H}_0(t) = [1-\lambda(t)] \mathcal{H}_\mathrm{d} + \lambda(t) \mathcal{H}_\mathrm{p},
\label{eq_H0}
\end{equation}
where $\lambda(t)$ is a time-dependent driving function that fulfills the boundary conditions $\lambda(t=0)=0$ and $\lambda(t=\tau)=1$ with $\tau$ the total annealing time. Reaching the exact ground state of $\mathcal{H}_\mathrm{p}$--- which for most interesting optimization problems is written in the form of single- and multi-spin $\sigma_i^z$ terms that describe high-order polynomial unconstrained binary optimization (PUBO) problems with $k$-local interactions--- generally requires adiabaticity, and the time necessary to satisfy this condition grows exponentially as a function of $N$ if the energy gap between the ground state and the first-excited state closes exponentially, which is the case in most of the interesting combinatorial optimization problems~\cite{albash2018adiabatic}.

\par 

To overcome this bottleneck, one can implement a so-called counter-diabatic  Hamiltonian $\mathcal{H}_\mathrm{CD}(t)$ to suppress transitions between the system eigenstates. The full Hamiltonian then reads
\begin{equation}
\mathcal{H}(t) = \mathcal{H}_0(t) + \mathcal{H}_\mathrm{CD}(t), 
\label{eq_H_full}
\end{equation}
where $\mathcal{H}_\mathrm{CD}(t)=\dot{\lambda}(t) \mathcal{A}_\lambda(t)$ is the additional counter-diabatic Hamiltonian, and $\mathcal{A}_\lambda(t) = i \hbar U^{\dagger}(t) \partial_{\lambda} U(t)$ with
\begin{equation}
    U(t) = {\cal T}\exp\left[-\frac{i}{\hbar}\int_0^{t}\mathcal{H}_0(t')dt'\right]
    \label{eq:unitary}
\end{equation}
is the exact time-dependent adiabatic gauge potential~\cite{sels2017minimizing, kolodrubetz2017geometry, claeys2019floquet} with respect to the driving function $\lambda(t)$ of Eq.~\eqref{eq_H0} and $\dot{\lambda}(t)$ its time derivative.

\par

Finding the exact adiabatic gauge potential is a challenging task and generally requires {\em a priori} knowledge of the system eigenstates for the whole annealing time as can be seen from the above expression of $\mathcal{A}_\lambda(t)$~\cite{berry2009transitionless}, which is impossible in practice. To overcome this difficulty, one can employ an approximate adiabatic gauge potential, denoted with a prime as $\mathcal{A}'_\lambda(t)$ (not to be confused with the derivative), which includes only local single-spin terms involving $\{ \sigma_i^y\}_i$ and which adds a new degree of freedom to the system~\cite{sels2017minimizing}. 
We note here that in the case of the original Hamiltonian $\mathcal{H}_0(t)$, given by Eq.~\eqref{eq_H0}, with driver Hamiltonian $\mathcal{H}_\mathrm{d}$, given by Eq.~\eqref{eq_H_driver}, and problem Hamiltonian $\mathcal{H}_\mathrm{p}$ including $\sigma^z$ terms, additional $\{ \sigma_i^x\}_i$ and $\{ \sigma_i^z\}_i$ operators do not yield further improvement (see Appendix~\ref{App_B} for the example of the Landau-Zener model for more details).

\par

Following the variational principle of Ref.~\cite{sels2017minimizing}, one finds the best possible approximate adiabatic gauge potential by defining a Hermitian operator $G_\lambda(\mathcal{A}_\lambda) \equiv \partial_\lambda \mathcal{H}_0 + i [\mathcal{A}_\lambda, \mathcal{H}_0]$ and minimizing the operator distance 
\begin{align}
    \mathcal{D}^2(\mathcal{A}'_\lambda)= \mathrm{Tr}\{[G_\lambda(\mathcal{A}_\lambda) - G_\lambda(\mathcal{A}'_\lambda)]^2\}
\end{align}
between the exact, $\mathcal{A}_\lambda$, and approximate, $\mathcal{A}'_\lambda$, adiabatic gauge potentials with respect to the parameters in $\mathcal{A}'_\lambda$.
This is equivalent to minimizing the action 
$\mathcal{S}(\mathcal{A}'_\lambda) = \mathrm{Tr}[G^2_\lambda(\mathcal{A}'_\lambda)]$
with respect to its parameters, symbolically written as $\delta \mathcal{S}(\mathcal{A}'_\lambda)/\delta \mathcal{A}'_\lambda=0$, as detailed in Refs.~\cite{sels2017minimizing, kolodrubetz2017geometry, claeys2019floquet} and Appendix \ref{App_A}.
In this general approach, the driver, $\mathcal{H}_\mathrm{d}$, and problem Hamiltonian $\mathcal{H}_\mathrm{p}$ are kept intact.

\par 

Now, we notice that the time-dependent coefficients of the two terms in the Hamiltonian $\mathcal{H}_0(t)$, given by Eq.~\eqref{eq_H0}, can be chosen independently--- not necessarily in a single-parameter form as in Eq.~\eqref{eq_H0}--- as long as the initial Hamiltonian is $\mathcal{H}_\mathrm{d}$ and the final Hamiltonian is $\mathcal{H}_\mathrm{p}$. We take advantage of this degree of freedom and choose to write these coefficients using \emph{two} time-dependent parameters $\lambda(t)$ and $\gamma(t)$ as
\begin{align}
\mathcal{H}^{\lambda, \gamma}_0(t) = -[1 - \lambda(t)]\gamma(t) \sum_{i=1}^N \sigma_i^x 
+ \lambda(t) \mathcal{H}_\mathrm{p},
\label{eq_generic_H}
\end{align}
where $\lambda (t)$ satisfies the same boundary conditions as before, $\lambda (0)=0$ and $\lambda (\tau) =1$, and $\gamma (t)$ is an arbitrary function satisfying $\gamma (0) \ne 0$ and which generalizes the form of $\gamma$ in Eq.~\eqref{eq_H_driver}.
Since we have an additional function $\gamma (t)$, it is natural to introduce a corresponding additional adiabatic gauge potential $\mathcal{A}_\gamma$. We therefore employ the approximate \emph{local} two-parameter CD Hamiltonian
\begin{equation}
\mathcal{H}^{\lambda, \gamma}_\mathrm{CD}(t)=\dot{\lambda}(t)\mathcal{A}'_\lambda(t) + \dot{\gamma}(t) \mathcal{A}'_\gamma(t),
\label{eq_H_CD_new} 
\end{equation}
where $\mathcal{A}'_\gamma(t)$ is also a linear combination of $\sigma_i^y$ but with a different coefficient than in $\mathcal{A}'_\lambda(t)$. 

\par 

One may wonder if the same linear combination of  $\sigma_i^y$ operators as in $\mathcal{A}'_\lambda(t)$ with just a different coefficient would lead to different results. As we will see in the next section, it indeed leads to an improvement of the annealing performance in several measures, thanks to the enhanced space of search for variational optimization of the coefficients as functions of time. See, also, Appendix~\ref{appendix:different_gamma}. 

As shown in Appendix~\ref{App_A}, finding the optimal coefficients in the two adiabatic gauge potentials $\mathcal{A}'_\lambda$ and $\mathcal{A}'_\gamma$ is equivalent to minimizing the two-parameter action,
\begin{equation}
\mathcal{S} = \mathrm{Tr}[G^2_\lambda(\mathcal{A}'_\lambda)] + \mathrm{Tr}[G^2_\gamma(\mathcal{A}'_\gamma)]
\label{eq_action}
\end{equation}
with respect to the parameters in the two adiabatic gauge potentials $\mathcal{A}'_\lambda$ and $\mathcal{A}'_\gamma$, i.e.,
\begin{align}
   \frac{\delta\mathcal{S}}{\delta \mathcal{A}'_\lambda}=0, \, \frac{\delta \mathcal{S}}{\delta \mathcal{A}'_\gamma} =0,
\end{align}
where $G_\gamma(\mathcal{A}'_\gamma) \equiv \partial_\gamma \mathcal{H}_0 + i [\mathcal{A}'_\gamma, \mathcal{H}_0]$ is the additional Hermitian operator with respect to $\gamma(t)$.

\par
 
As we will see later, the introduction of an additional time dependence for the transverse magnetic field strength, $\gamma(t)$, and thus the emergence of the additional adiabatic gauge potential $\mathcal{A}'_\gamma(t)$ has significant consequences for local CD driving. The operator distance and thus the corresponding action, given by Eq.~\eqref{eq_action}, can be algebraically determined for a given set of two functions $\lambda(t)$ and $\gamma(t)$.  A detailed derivation of Eqs.~\eqref{eq_H_CD_new} and \eqref{eq_action} can be found in Appendix~\ref{App_A} and its application on the easy single-body Landau-Zener model in Appendix~\ref{App_B}.

Although it is desirable to find the best possible functional forms of $\lambda(t)$ and $\gamma(t)$, this poses an additional complex step of functional optimization, about which we do not have a clear principle to rely upon.  Indeed, existing studies adopt simple functions satisfying boundary conditions without elaborating on further optimization of functional forms \cite{sels2017minimizing, kolodrubetz2017geometry, claeys2019floquet} . We follow this tradition and work with simple conventional forms of those functions as illustrated in the next section and delegate the optimization of those functions to a future project. See Appendix \ref{appendix:different_gamma} for additional information.

\subsection*{\pmb{$p$}-spin model}
\begin{figure*}[htbp]
  \includegraphics[width=.99\textwidth]{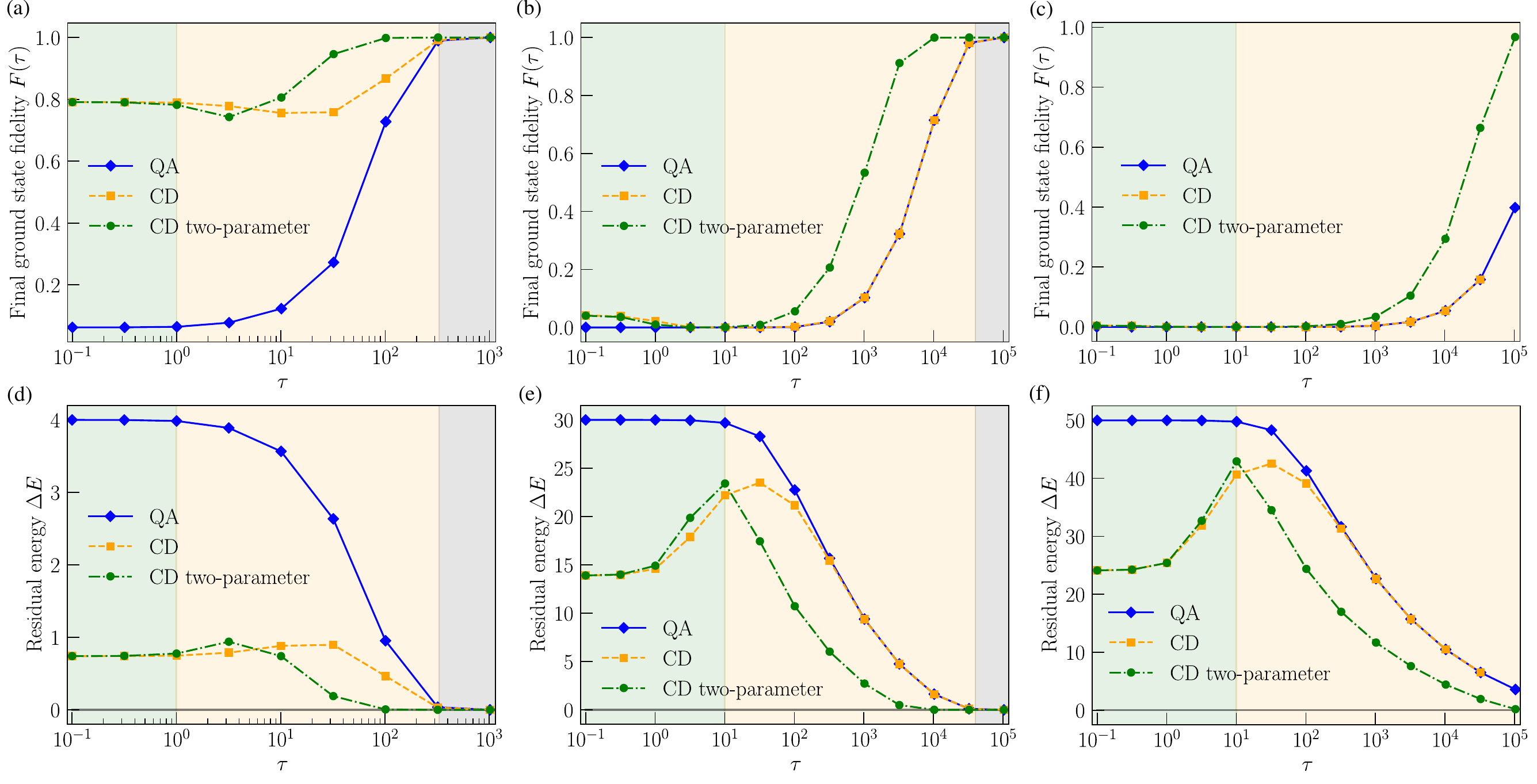}
  \caption{\textbf{Ground-state fidelity and residual energy.} (a)--(c)~Final ground-state fidelity and (d)--(f)~residual energy for (i)~traditional quantum annealing (diamonds, blue solid line), (ii)~single-parameter CD drive (squares, orange dashed line), and (iii)~two-parameter CD drive (circles, green dash-dotted line) as functions of annealing time $\tau$. The system sizes are (a),(d) $N=4$, (b),(e) $N=30$, and (c),(f) $N=50$, where $\gamma_\mathrm{init}=0.1$ for all panels. Time ranges are color coded as follows: short-time regime (green-shaded areas) where the fidelity is approximately $1/2^N$ for traditional quantum annealing, long-time regime (yellow-shaded areas) where transient behavior is observed and the two-parameter CD drive shows a clear advantage, and adiabatic regime (gray-shaded areas) where $F(\tau) > 0.99$.}
  \label{fig:p_spin}
\end{figure*}
Our method can be applied to any problem Hamiltonian $\mathcal{H}_\mathrm{p}$. In the present paper, we test our approach by using the $p$-spin model with $p=3$ as the problem Hamiltonian since it is a hard problem for traditional quantum annealing due to a first-order quantum phase transition, though the final ground state is trivially known to be ferromagnetic~\cite{joerg2010energy, seki2012quantum, seki2015quantum, seoane2012manybody, nishimori2017exponential}. Another advantage of the $p$-spin model is that the total spin quantum number is conserved, 
which can also be stated as that the Hamiltonian is invariant under an arbitrary permutation of site indices.  This fact makes it possible to study very large system sizes numerically by restricting ourselves to the subspace of a fixed spin quantum number corresponding to the ground state, as we shall see in the next section. 

\par 

The total Hamiltonian of interest reads
\begin{align}
\mathcal{H}^{\lambda, \gamma}_0(t) = -[1 - \lambda(t)] \gamma(t)\sum_{i=1}^N \sigma_i^x 
- \lambda(t) N \left( \frac{1}{N} \sum^N_{i=1} \sigma^z_i \right)^3.
\label{eq_H0_p_spin}
\end{align}
This Hamiltonian fulfills the commutation relation
\begin{align}
    [\mathcal{H}^{\lambda, \gamma}_0(t), {\bf S}_{\rm total}^2]=0,
\end{align}
where ${\bf S}_{\rm total}=(S_{\rm total}^x, S_{\rm total}^y, S_{\rm total}^z)$, is the total spin quantum number with \mbox{$S_{\rm total}^x=(1/2)\sum_{i=1}^N \sigma_i^x$} and similarly for the $y$ and $z$ components. Since the initial condition is that the ground state of $\mathcal{H}_\mathrm{d}$ of Eq.~(\ref{eq_H_driver}) is an eigenstate of ${\bf S}_{\rm total}^2$ with largest eigenvalue, we can restrict our numerical computations to the space of this eigenvalue, which greatly reduces the dimension of the Hilbert space to be explored numerically from exponential to linear in $N$.

\par 

Throughout this work, we will use the driving functions
\begin{align}
\lambda(t) &= \sin^2\left[\dfrac{\pi}{2}\sin^2\left(\dfrac{\pi t}{2 \tau}\right)\right], \nonumber \\
\gamma(t) &= \gamma_\mathrm{init} + \lambda(t),
\label{eq_driving_functions}
\end{align}
where we have chosen the function $\lambda (t)$ following Ref.~\cite{sels2017minimizing}. The above form of $\gamma(t)$ is chosen arbitrarily and its deeper investigation is a future task as mentioned before. We note here that $\gamma(t)$ can generally take a most generic form as long as $\gamma(t)\ne 0$ and does not necessarily need to include $\lambda(t)$. We have checked numerically that small variations of the value of $\gamma_\mathrm{init}$ do not lead to noticeable changes of the results.  

\par 

Since any local adiabatic gauge potential is a linear combination of $\sigma_i^y$ (\cf Appendix~\ref{App_B}), we write, for the latter,
\begin{align}
    \mathcal{A}'_\lambda = \sum_{i=1}^N \alpha \sigma_i^y,\quad\mathcal{A}'_\gamma= \sum_{i=1}^N \beta \sigma_i^y.
    \label{eq_AGPs}
\end{align}
We choose $\alpha$ and $\beta$ to be site independent reflecting the permutation symmetry of the $p$-spin Hamiltonian of Eq.~(\ref{eq_H0_p_spin}). These coefficients can generally be chosen to depend on the site index $i$ for problems without such symmetries.

Minimizing the corresponding action $\mathcal{S}$, given by Eq.~(\ref{eq_action}), with respect to the coefficients $\alpha$ and $\beta$, as detailed in Appendix \ref{App_C}, 
leads to their optimal algebraic solutions and thus the CD Hamiltonian, given by Eq.~\eqref{eq_H_CD_new}, as
\begin{align}
&\mathcal{H}^{\lambda, \gamma}_\mathrm{CD}(t)=\sum_{i=1}^N (\dot{\lambda} \alpha + \dot{\gamma} \beta) \sigma_i^y, \nonumber \\
    &\alpha = - \kappa \gamma , \: \: \:\: \beta = \kappa (1-\lambda)\lambda, \nonumber \\
      &\kappa = \frac{1}{2}\frac{ N^2(3N-2)}{ (1-\lambda)^2\gamma^2N^4 + \lambda^2 (27 N^2 - 66 N +40) }.
    \label{eq_solution_p_spin}
\end{align}
It is noticed that $\kappa$ is proportional to $1/N$ for large $N$ and thus the CD Hamiltonian $\mathcal{H}^{\lambda, \gamma}_\mathrm{CD}(t)$ becomes small for very large $N$ \footnote{It is known that in the $p$-spin model the thermodynamic limit $N\to\infty$ is equivalent to the classical limit and a correction of $\mathcal{O}(1/N)$ is equivalent to a quantum correction of order $\mathcal{O}(\hbar)$~\cite{ohkuwa2017exact}. Thus the present adiabatic gauge potential may be regarded as representing delicate quantum effects in the $p$-spin model.}. We therefore expect that the effect of the adiabatic gauge potentials is seen most prominently for relatively small to moderate $N$. This also means that, as long as the $p$-spin model is concerned, the present method does not lead to a drastic scaling advantage that reduces the asymptotic computational complexity from exponential to polynomial in the limit of very large $N$, although significant improvements will be observed numerically even for moderately large $N$, as we will see in the next section. 

The corresponding full Hamiltonian then reads
\begin{align}
   \mathcal{H}^{\lambda, \gamma}(t) = \mathcal{H}^{\lambda, \gamma}_0(t) + \sum_{i=1}^N (\dot{\lambda} \alpha + \dot{\gamma} \beta) \sigma_i^y \label{eq:H0_plus_y}
\end{align}
with the solutions $\alpha$ and $\beta$, given by Eq.~\eqref{eq_solution_p_spin}, and $\dot{\lambda}$ and $\dot{\gamma}$ the time derivatives of Eq.~\eqref{eq_driving_functions}.

\par 

To facilitate experimental implementation, we eliminate the $\sigma_i^y$ terms by rotating this full Hamiltonian around the $z$ axis in spin space, i.e., applying the unitary gauge transformation 
\begin{align}
    U_\mathrm{g}(t) = \exp\left[i \frac{\theta(t)}{2} \sum_i \sigma_i^z \right]
    \label{eq:unitary_transformation}
\end{align}
over the angle $\theta(t) = \arctan (Y / X)$, with $X = - (1 - \lambda) \gamma$ and $Y = \dot{\lambda} \alpha + \dot{\gamma} \beta$. 
The resulting effective Hamiltonian in the laboratory frame then has the form
\begin{align}
\mathcal{H}^{\lambda, \gamma}_\mathrm{eff}(t) &= \sum_{i=1}^N \sqrt{X^2 + Y^2} \sigma_i^x - \lambda(t) \frac{6}{N^2}  \sum^N_{i< j< k} \sigma^z_i \sigma^z_j \sigma^z_k \nonumber \\
&- \sum_{i=1}^N \left[\dfrac{1}{2}\frac{X\dot{Y} - Y \dot{X}}{X^2 + Y^2} + \lambda(t) \dfrac{3N-2}{N^2} \right] \sigma_i^z
\label{eq_H_full_eff}
\end{align}
(see Appendix~\ref{App_C} and Ref.~\cite{sels2017minimizing} for additional details). This Hamiltonian consists only of $\sigma_i^x$ and $\sigma_i^z$ terms, which makes it more feasible for experimental realization than Eq.~\eqref{eq:H0_plus_y} with $\sigma_i^y$.

\section{Numerical verification}\label{sec_results}
We next present numerical results of our method for the $p$-spin model with $p=3$. To this end, we compute the final ground-state fidelity $F(\tau)=|\langle \psi(\tau) | \phi_0 \rangle|^2$, with $|\psi(\tau) \rangle$ and $|\phi_0 \rangle$ the states at the end of annealing and the true ground state of the problem Hamiltonian, respectively, and residual energy $\Delta E = E(\tau) - E_0$, with $E(\tau)$ and $E_0$ being the energy at the end of annealing and the true ground-state energy, respectively. 
We compare three protocols: (i)~traditional quantum annealing with the original Hamiltonian [$\mathcal{H}^{\lambda, \gamma}_0(t)$, given by Eq.~\eqref{eq_H0_p_spin}], (ii)~the existing method with single-parameter CD Hamiltonian [Eq.~\eqref{eq_H_full_eff} with $\gamma(t)=\gamma_\mathrm{init}$ and thus $\beta=0$] and (iii)~two-parameter CD Hamiltonian [$\mathcal{H}^{\lambda, \gamma}_\mathrm{eff}(t)$, Eq.~\eqref{eq_H_full_eff}]. We test a wide range of annealing times $\tau$ from $10^{-1}$ to $10^5$ and different system sizes up to $N=100$ by exploiting the spin symmetry of the problem.

\par 

We numerically solved the Schr\"odinger equation for the Hamiltonian dynamics and computed the fidelity, residual energy, and the time-to-solution, which is a measure of the effective annealing time to reach the solution of the optimization problem with probability $p_\mathrm{r}$~\cite{albash2018demonstration}, i.e.,
\begin{equation} 
\mathrm{TTS}(\tau) = \begin{cases}
\tau \,\dfrac{\ln(1-p_\mathrm{r})}{\ln[1-F(\tau)]} & \, \textrm{for} \, F(\tau)<1, \\
\; \; \; \; \; \; \; \; \tau & \, \textrm{for} \, F(\tau)=1,
\end{cases}
\label{eq_TTS}
\end{equation}
where we have set $p_\mathrm{r}=0.99$ as the success probability threshold.
For our numerical computations, we used QUTIP 4.5~\cite{johannson2013qutip}.

\begin{figure*}[ht]
  \centering
  \includegraphics[width=.75\textwidth]{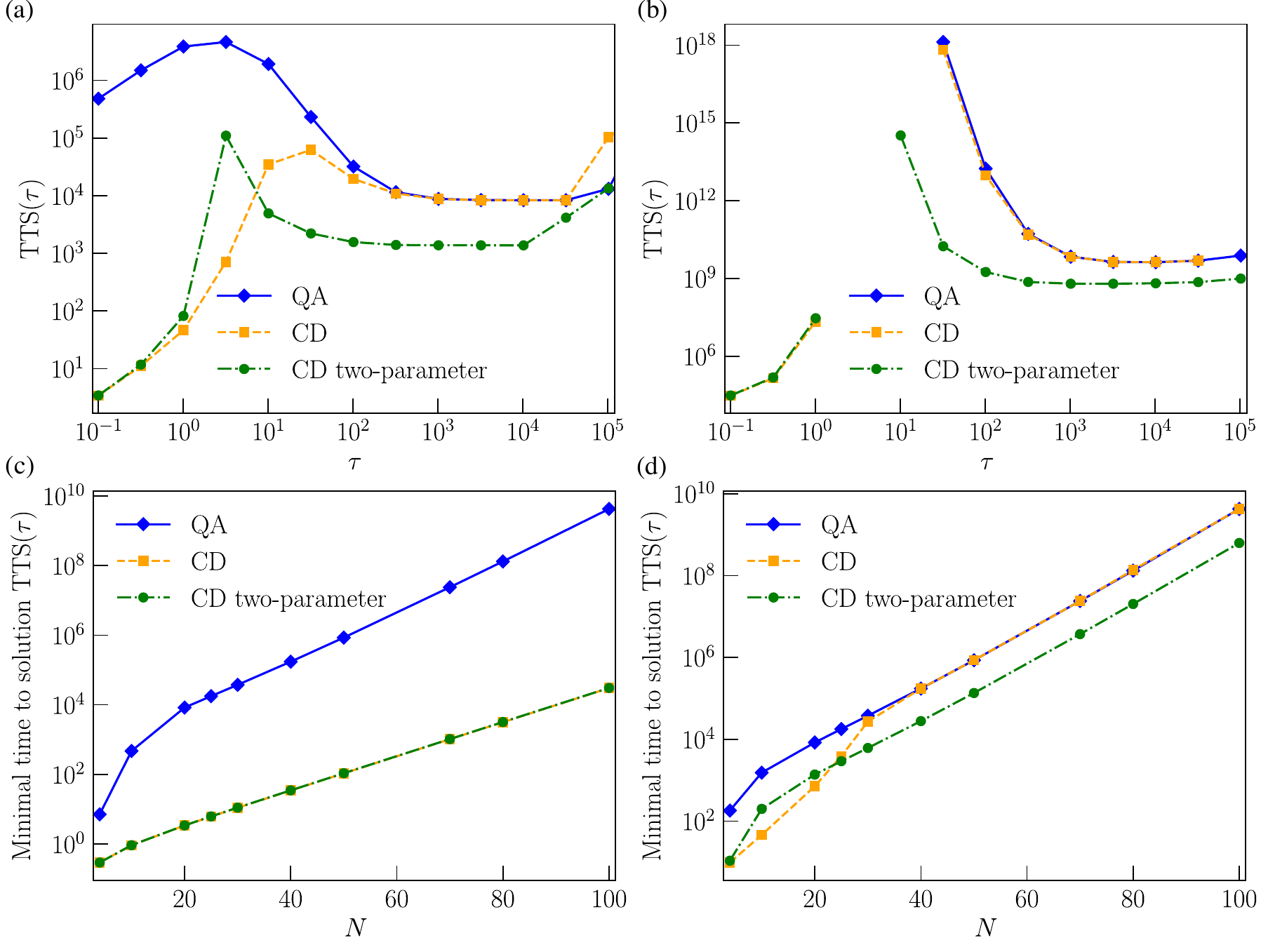}
  \caption{\textbf{Time-to-solution.} Time-to-solution $\mathrm{TTS}(\tau)$ for (i)~traditional quantum annealing (diamonds, blue solid line), (ii)~single-parameter CD drive (squares, orange dashed line), and (iii)~two-parameter CD drive (circles, green dash-dotted line) for (a)~$N=20$ and (b)~$N=100$. For the latter, the data between $\tau=1$ to about 10 are not shown because the values are too large to achieve reasonable numerical precision.  Minimal time-to-solution for (c)~short-time region ($\tau \lesssim 1$) and (d)~long-time region ($\tau \gtrsim 10$). Other parameters are the same as in Fig.~\ref{fig:p_spin}.}
  \label{fig:scaling}
\end{figure*}

\subsection*{Dependence on annealing time}
Figure~\ref{fig:p_spin} depicts the final ground-state fidelity $F(\tau)$ [Figs.~\ref{fig:p_spin}\textcolor{red}{(a)}--\ref{fig:p_spin}\textcolor{red}{(c)}] and residual energy $\Delta E$ [Figs.~\ref{fig:p_spin}\textcolor{red}{(d)}--\ref{fig:p_spin}\textcolor{red}{(f)}] as functions of annealing time $\tau$ for system sizes $N=4$ [Figs.~\ref{fig:p_spin}\textcolor{red}{(a)} and~\ref{fig:p_spin}\textcolor{red}{(d)}], $N=30$ [Figs.~\ref{fig:p_spin}\textcolor{red}{(b)} and~\ref{fig:p_spin}\textcolor{red}{(e)}] and $N=50$ [Figs.~\ref{fig:p_spin}\textcolor{red}{(c)} and~\ref{fig:p_spin}\textcolor{red}{(f)}].

For the original Hamiltonian without the CD term [$\mathcal{H}_0(t)$, given by Eq.~\eqref{eq_H0_p_spin}; diamond with blue solid line in the figure], we see that the final state is far away from the ground state for short annealing time $\tau$ [green-shaded areas, where $F(\tau) \approx 1/2^N$ for $\mathcal{H}_0(t)$] and reach the final ground state in the adiabatic regime [gray-shaded areas, where $F(\tau) > 0.99$] for very long annealing time.

The existing method of a single-parameter CD driven Hamiltonian [Eq.~\eqref{eq_solution_p_spin} with $\gamma(t)=\gamma_\mathrm{init}$ and thus $\beta=0$; square, orange dashed line] reaches a considerably higher final ground-state fidelity and lower residual energy, respectively, especially for short annealing time (green-shaded areas), yet approaches their original counterpart for longer annealing time (yellow-shaded areas) due to the fact that $\dot{\lambda} \propto 1/\tau$ [\cf Eq.~\eqref{eq_driving_functions}]. Consequently, the counter-diabatic Hamiltonian $\mathcal{H}^{\lambda,\gamma}_\mathrm{CD}(t)$ naturally converges towards zero for longer annealing time, in particular in the adiabatic limit (gray-shaded areas), and thus does not yield any further speedup.

On the other hand, for the two-parameter CD driven Hamiltonian [$\mathcal{H}^{\lambda, \gamma}_\mathrm{CD}(t)$, given by Eq.~\eqref{eq_H_full_eff}, where $\gamma(t)=\gamma_\mathrm{init} + \lambda(t)$; circle, green dash-dotted line], it is observed that we reach considerably higher final ground-state fidelity and lower residual energy compared to traditional quantum annealing (QA) and single-parameter CD driving for the long annealing time regime (yellow-shaded areas). This is important since the asymptotic adiabatic regime (gray-shaded areas) starts at later times for larger system sizes as seen in Fig.~\ref{fig:p_spin}\textcolor{red}{(c)}, meaning that the system performance in the long-, but not yet adiabatic, time regime (yellow-shaded areas) becomes more and more vital for larger systems. 
In other words, for the two-parameter CD driven Hamiltonian, we come closer to the adiabatic regime more quickly, thus performing much better (around an order of magnitude reduction in annealing time to reach the same values of fidelity and residual energy) compared to its traditional quantum annealing and single-parameter CD driving counterparts.
Although the last term in Eq.~(\ref{eq:H0_plus_y}) with $\dot{\gamma}\beta$ may superficially seem not to add a new element to the single-parameter method just with $\dot{\lambda}\alpha$, the present numerical results clearly indicate that our two-parameter method leads to significant advantages in the intermediate-time region (yellow-shaded region in Fig.~\ref{fig:p_spin}).  This time region is important in practice because, first, the gray-shaded adiabatic region is often hard to reach for very large systems, and, second, the green shaded short-time region has large residual energy and low fidelity.

\subsection*{Time-to-solution}
We further studied the time-to-solution, a central measure of annealing time necessary to reach the solution of the optimization problem of interest with a certain high success probability, for different system sizes $N$. 
Figures~\ref{fig:scaling}\textcolor{red}{(a)} and~\ref{fig:scaling}\textcolor{red}{(b)} depict the time-to-solution $\mathrm{TTS}(\tau)$, given by Eq.~\eqref{eq_TTS}, for fixed system sizes $N=20$ and $N=100$, respectively. It is observed that the minimal time-to-solution is located at the shortest annealing time that we studied, i.e. $\tau=10^{-1}$, except for the case of traditional quantum annealing. We did not study even shorter time ranges because the time derivative of $\lambda(t)$, given by Eq.~\eqref{eq_driving_functions}, becomes anomalous for very small $\tau$ and also experimental implementation may be difficult for too short annealing time. We further found a local minimum of $\mathrm{TTS}(\tau)$ at a longer time, $\tau \approx 10^3$. 

\par

Figure~\ref{fig:scaling}\textcolor{red}{(c)} depicts the system size dependence of the minimal time-to-solution at the shortest annealing time that we studied, $\tau=10^{-1}$. We see that the existing single-parameter CD method and our two-parameter method have a scaling advantage over traditional quantum annealing in the sense that the slope is smaller, i.e., a smaller constant in the exponent. Figure~\ref{fig:scaling}\textcolor{red}{(d)} depicts the time-to-solution at the local minimum $\tau \approx 10^3$ as a function of the system size. Our two-parameter approach has the same scaling behavior (the same slope) as the other two methods but depicts a constant speedup of the order of around 10. The same scaling behavior for large $N$ is not very surprising because the adiabatic gauge potentials $\mathcal{A}'_\lambda$ and $\mathcal{A}'_\gamma$ are proportional to $1/N$ and will consequently disappear for increasing system sizes. We notice here that this asymptotic vanishing of the adiabatic gauge potentials is a special property of the $p$-spin model, and the advantage of the present method is expected to remain finite for large system size and large annealing time in other models, for which we have preliminary analytical and numerical evidence. The comparison of Figs.~\ref{fig:scaling}\textcolor{red}{(c)} and~\ref{fig:scaling}\textcolor{red}{(d)} reveals that it is more advantageous to repeat very short annealing processes many times than to run a single long annealing, at least in the present problem.

\par 

Our preliminary data for a few other problem Hamiltonians indicate the possibility that the absolute minimum at the shortest annealing time may be a finite-size effect and seems to vanish for large system sizes and, in particular, in the thermodynamic limit. If this proves to be true, the $p$-spin model is peculiar in the sense that finite-size effects persist even for system sizes as large as $N=100$. Whether or not this behavior is shared by other problem Hamiltonians is an interesting future topic of research.  

\subsection*{Behavior of coefficients}
\begin{figure}
  \centering
  \includegraphics[width=.95\columnwidth]{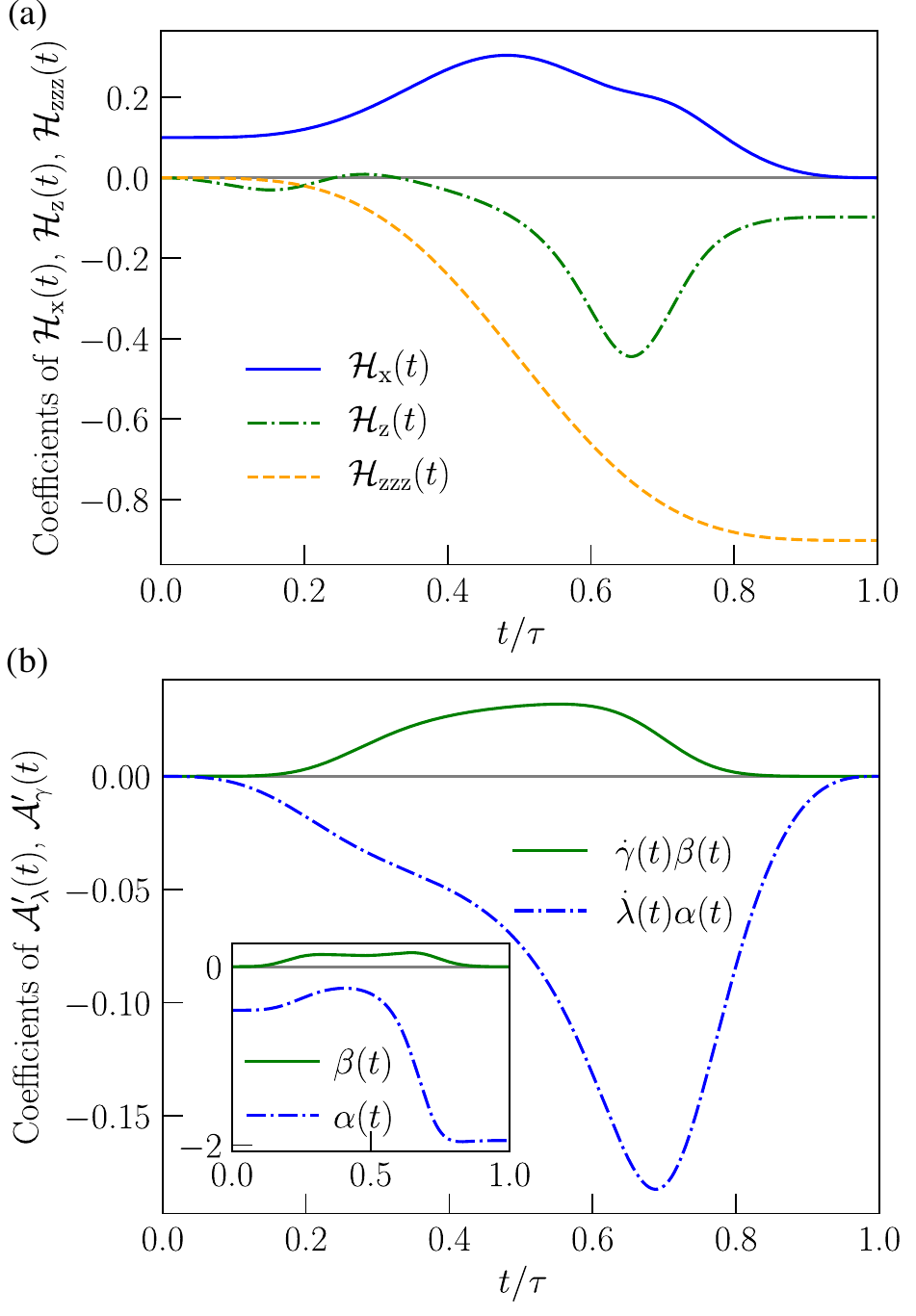}
  \caption{\textbf{Coefficients of two-parameter CD Hamiltonian.} (a)~Time dependence of coefficients of terms of the Hamiltonian as described in the text, $\mathcal{H}_\mathrm{x}(t)$ (upper blue solid line), $\mathcal{H}_\mathrm{z}(t)$ (middle green dash-dotted line) and $\mathcal{H}_\mathrm{zzz}(t)$ (lower orange dashed line), the last one being scaled by $N$ to fairly compare coefficients of extensive operator terms.  (b)~Time dependence of the coefficients of the adiabatic gauge potentials $\mathcal{A}'_\lambda(t)$ (lower blue dash-dotted line) and $\mathcal{A}'_\gamma(t)$ (upper green solid line). Inset depicts the corresponding coefficients $\alpha(t)$ and $\beta(t)$ under the two-parameter CD drive with $\gamma(t)=0.1 + \lambda(t)$, annealing time $\tau=10$ and system size $N=30$. Other parameters are the same as in Fig.~\ref{fig:p_spin}.}
  \label{fig:coefficients_p_spin}
\end{figure}

Figure~\ref{fig:coefficients_p_spin}\textcolor{red}{(a)} depicts the time dependence of the coefficient of each term of the full Hamiltonian $\mathcal{H}^{\lambda, \gamma}_\mathrm{eff}(t)$, given by Eq.~\eqref{eq_H_full_eff}, in the rotated frame, i.e.,
\begin{align}
   &\mathcal{H}_\mathrm{x}(t)=\sum_{i=1}^N \sqrt{X^2 + Y^2}\sigma_i^x, \label{eq_Hx} \\
   &\mathcal{H}_\mathrm{zzz}(t)=-\lambda(t)\frac{6}{N^2} \sum_{i<j<k}  \sigma_i^z \sigma_j^z \sigma_k^z, \label{eq_Hzzz} \\
   &\mathcal{H}_\mathrm{z}(t)=- \sum_{i=1}^N \left[\frac{1}{2}\frac{X \dot{Y} - \dot{X} Y}{X^2 + Y^2} + \lambda(t) \frac{3N-2}{N^2}\right] \sigma_i^z, \label{eq_Hz}
\end{align}
for annealing time $\tau=10$, system size $N=30$, and other parameters as in Fig.~\ref{fig:p_spin}. Figure~\ref{fig:coefficients_p_spin}\textcolor{red}{(b)} depicts the coefficients of the adiabatic gauge potentials $\mathcal{A}'_\lambda(t)$ and $\mathcal{A}'_\gamma(t)$ and the corresponding coefficients $\alpha(t)$ and $\beta(t)$, given by Eq.~\eqref{eq_solution_p_spin}, in the inset.
The maximal corresponding strengths of the additional magnetic field in the $y$ direction in the original frame [Fig.~\ref{fig:coefficients_p_spin}\textcolor{red}{(b)}] and in the rotated frame [reflected in the coefficients of $\mathcal{H}_\mathrm{x}(t)$ and $\mathcal{H}_\mathrm{z}(t)$ in Fig.~\ref{fig:coefficients_p_spin}\textcolor{red}{(a)}] are not (overwhelmingly) larger than the original parameters in $\mathcal{H}_\mathrm{zzz}(t)$ for this annealing time regime, which makes this approach attractive for experimental realization.

\subsection*{Energy spectrum of two-parameter CD drive}
It is useful to see how the wave function is spread over the instantaneous eigenstates of the rotated full Hamiltonian, given by Eq.~\eqref{eq_H_full_eff}, during the present two-parameter CD drive in the laboratory frame.
\begin{figure*}[htb]
  \centering
  \includegraphics[width=.99\textwidth]{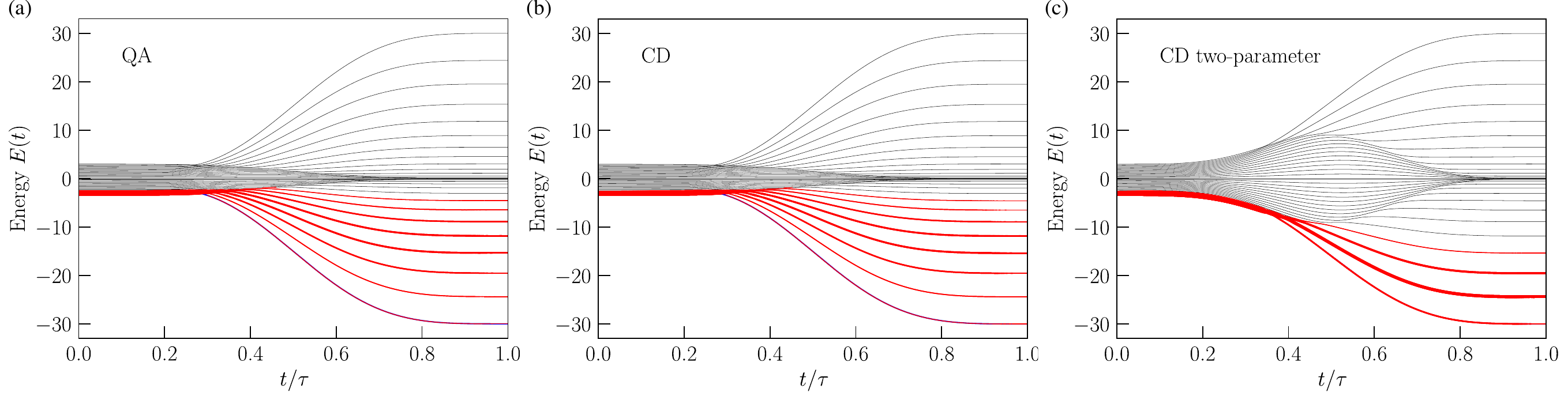}
  \caption{\textbf{Energy spectrum.} Instantaneous energy spectrum $E(t)$ for (a)~traditional quantum annealing, (b)~single-parameter CD driving and (c)~two-parameter CD driving for system size $N=30$, annealing time $\tau=300$ and $\gamma_\mathrm{init}=0.1$. Thickness of red curves indicates the occupation probability of each eigenstate in the dynamical processes of the three annealing protocols. Higher excited states have a neglectably small occupation probability and thus the corresponding very thin red lines can not be seen.}
  \label{fig:energyspectrum}
\end{figure*}

Figure~\ref{fig:energyspectrum} depicts the occupation probability of each instantaneous eigenstate, expressed by the thickness of the red lines, for a system size $N=30$ and annealing time $\tau=300$, corresponding to Figs.~\ref{fig:p_spin}\textcolor{red}{(b)} and~\ref{fig:p_spin}\textcolor{red}{(e)}, where the two-parameter CD drive shows a clear advantage over traditional QA and the existing method of single-parameter CD driving.
We observe that Figs.~\ref{fig:energyspectrum}\textcolor{red}{(a)} and~\ref{fig:p_spin}\textcolor{red}{(b)} share a very similar eigenspectrum, and the wavefunction is spread over many excited states after $t/\tau \approx 0.3$ via a cascade of avoided level crossings. In contrast, in the two-parameter CD case [Fig.~\ref{fig:p_spin}\textcolor{red}{(c)}] the structure of the eigenspectrum has significantly changed and the system is driven downward in the spectrum around $t/\tau \approx 0.3$, which results in the high occupation probabilities in low-energy eigenstates in the end of the annealing process. We emphasize that such an ingenious protocol has emerged naturally from the two-parameter variational approach to suppress undesirable diabatic transitions observed in Figs.~\ref{fig:p_spin}\textcolor{red}{(a)} and~\ref{fig:p_spin}\textcolor{red}{(b)}. 

\section{Discussion and conclusion}\label{sec_discussion}
We have proposed and tested a method to find an efficient local CD Hamiltonian that outperforms its traditional quantum annealing and single-parameter approximate CD counterparts with respect to enhanced final ground-state fidelity and reduced residual energy as well as time-to-solution. The method introduces an additional term in the adiabatic gauge potential by taking advantage of the degree of freedom of choosing a time-dependent transverse magnetic field strength. 
This two-parameter local CD approach generalizes the existing method of single-parameter CD driving by expanding the search space of optimal parameters by introducing a second controllable driving function $\gamma(t)$.
The corresponding CD Hamiltonian in this approach is local and can be expressed, after a rotation in spin space, just in terms of the usual transverse-field Ising model but with unconventional diabatic control of the magnetic field strengths. For the goal of performance improvement, the latter may thus be implemented in current quantum annealing devices on various platforms considerably more easily than other approaches, which introduce more involved terms into the Hamiltonian such as two-body $\sigma_i^x \sigma_j^x$ interactions.

\par 

We have tested the idea using the $p$-spin model with $p=3$ because it is possible to simulate the Schr\"odinger dynamics numerically for very large system sizes for this model due to its special symmetry of conserved total quantum spin number. We have derived the algebraic expression of the two-parameter CD Hamiltonian and numerically demonstrated a considerable increase in final ground-state fidelity and reduction in residual energy as well as time-to-solution compared to traditional quantum annealing and the single-parameter CD Hamiltonian approach. We further demonstrated a scaling advantage of time-to-solution of the approximate single- and two-parameter CD methods in the short-time region, and a constant speedup of the two-parameter method in the long-time region. 
The division of annealing processes in short-time and long-time regions has important numerical and operational consequences. Whereas the time-to-solution in the short-time region depicts a global minimum, the experimental realization of the strongly increasing additional magnet fields in this time region constitutes a severe hindrance for practical purposes. The local minimum of time-to-solution in the long-time region serves as a promising regime for experimental implementation since the additional magnetic fields are not (considerably) larger than their original analogs.
The lack of scaling advantage in the long-time region may originate in the $1/N$ scaling of the coefficients of the CD Hamiltonian for the $p$-spin model, which is a special property of this multi-body mean-field-like problem. We may expect an even better scaling behavior in many other problems where those coefficients of the CD Hamiltonian generally stay finite in the large-$N$ limit. 
Even when a clear scaling advantage is not achieved, the present method becomes useful at least for a quantitative improvement as exemplified in the $p$-spin model. In particular, our method may be realized in an improvement of existing annealing devices by a better control of system parameters of the transverse-field Ising model only. We note that the method can also be applied in the case of additional random longitudinal magnetic fields where site-dependent optimal algebraic solutions for the coefficients of the adiabatic gauge potentials can be easily found.
As a consequence, it does not need further additional terms to be realized experimentally and is versatile to be applicable to any problem, in contrast to other approaches such as non-stoquastic catalysts~\cite{seki2012quantum, seoane2012manybody, seki2015quantum, nishimori2017exponential} and inhomogeneous field driving~\cite{susa2018exponential, susa2018quantum, hartmann2019quantum},
in which one should determine in advance if the idea works in a given problem and, if it does, should find a proper way to meticulously control the system parameters, which is in general highly non-trivial for a generic optimization problem. 

\par 

We have also illustrated how the two-parameter CD Hamiltonian resolves the problem of excitation to higher-energy states by showing the modification of the energy eigenspectrum that eliminates a cascade of avoided level crossings toward higher-energy states. It is an interesting future problem to identify problems in which this mechanism leads to a clear scaling advantage even for very large system sizes. Such examples may well exist because of the special disadvantageous property of the $p$-spin model as described above, i.e., that the coefficients of the CD Hamiltonian tends to vanish for larger system size.

\par 

We note that there exist other approaches to optimize the time dependence of coefficients in quantum annealing, e.g., from the viewpoint of optimal control theory and related ideas often under the context of the quantum approximate optimization algorithm \cite{yang2017optimizing, mbeng2019quantum, brady2021optimal, zhou2020experimental} (see, also, Ref.~\cite{takahashi2017shortcuts} for a related idea of inverse engineering). It is not clear {\em a priori} whether or not our two-parameter CD Hamiltonian is better in comparison with these approaches since the criteria of optimality are different. The comparison in terms of relevant physical quantities such as fidelity, residual energy and the time-to-solution will be the best way to measure the performance of different protocols.  It can happen that one is better than the other in some problems and the reverse in other problems, which reveals an interesting future topic to be studied.

\begin{acknowledgments}
We thank Kazutaka Tahakashi for useful comments. This work was supported by the Austrian Science Fund (FWF) through a START grant under Project No. Y1067-N27 and the SFB BeyondC Project No. F7108-N38, the Hauser-Raspe Foundation, and the European Union's Horizon 2020 research and innovation program under Grant Agreement No. 817482. This material is based upon work supported by the Defense Advanced Research Projects Agency (DARPA) under Contract No. HR001120C0068. Any opinions, findings and conclusions or recommendations expressed in this material are those of the author(s) and do not necessarily reflect the views of DARPA. The research is also based upon work partially supported by the Office of the Director of National Intelligence (ODNI), Intelligence Advanced Research Projects Activity (IARPA) and the Defense Advanced Research Projects Agency (DARPA), via the U.S. Army Research Office Contract No. W911NF-17-C-0050. The views and conclusions contained herein are those of the authors and should not be interpreted as necessarily representing the official policies or endorsements, either expressed or implied, of the ODNI, IARPA, DARPA, or the U.S. Government. The U.S. Government is authorized to reproduce and distribute reprints for Governmental purposes notwithstanding any copyright annotation thereon.
\end{acknowledgments}

\appendix
\newpage

\section{Derivation of adiabatic gauge potentials}\label{App_A}
In this appendix, we derive the two adiabatic gauge potentials $\mathcal{A}_\lambda$ and $\mathcal{A}_\gamma$ by considering a quantum state $|\psi\rangle$ evolving under the time-dependent Hamiltonian $\mathcal{H}^{\lambda, \gamma}_0(t)$.

The effective Schr\"odinger equation $i\partial_t |\psi\rangle = \mathcal{H}^{\lambda, \gamma}_0 |\psi
 \rangle$ in the moving frame by applying the unitary transformation $U=U(\lambda, \gamma)$ with $|\psi(t)\rangle = U^\dag|\psi\rangle$, and thus $|\psi\rangle = U|\psi(t)\rangle$, is written as
\begin{align}
&i\partial_t [U|\psi(t)\rangle] = \mathcal{H}^{\lambda, \gamma}_0[U|\psi(t)\rangle] \nonumber \\
&i (\partial_{\lambda} U \dot\lambda + \partial_{\gamma} U \dot\gamma)|\psi(t)\rangle + i U \partial_t|\psi(t)\rangle  = \mathcal{H}^{\lambda, \gamma}_0U|\psi(t)\rangle.
\end{align}
If we apply $U^\dag$ from the left, we have
\begin{align}
&i(U^\dag\partial_{\lambda} U \dot\lambda + U^\dag\partial_{\gamma} U \dot\gamma)|\psi(t)\rangle + i U^\dag U \partial_t|\psi(t)\rangle \nonumber\\
&= U^\dag\mathcal{H}^{\lambda, \gamma}_0U|\psi(t)\rangle
\end{align}
and consequently
\begin{align}
i\partial_t |\psi(t)\rangle  = \tilde{\mathcal{H}}^{\lambda, \gamma}_0|\psi(t)\rangle - i(U^\dag\partial_{\lambda} U \dot\lambda + U^\dag\partial_{\gamma} U \dot\gamma)|\psi(t)\rangle
\end{align}
which we write as
\begin{align}
i\partial_t |\psi(t)\rangle  = \tilde{\mathcal{H}}^{\lambda, \gamma}_0|\psi(t)\rangle - (\dot\lambda \mathcal{\tilde{A}}_{\lambda} + \dot\gamma \mathcal{\tilde{A}}_{\gamma})|\psi(t)\rangle,
\end{align}
where $\tilde{\mathcal{H}}^{\lambda, \gamma}_0(\lambda, \gamma)=U^\dagger \mathcal{H}^{\lambda, \gamma}_0 U$ is diagonal in its instantaneous eigenbasis, and $\mathcal{\tilde{A}}_{\lambda} = iU^\dag \partial_{\lambda} U$ and $\mathcal{\tilde{A}}_{\gamma} = iU^\dag \partial_{\gamma} U$ are the corresponding adiabatic gauge potentials in the moving frame with respect to the two time-dependent driving parameters $\lambda(t)$ and $\gamma(t)$, respectively.

\par 

The counter-diabatic Hamiltonian with respect to these two adiabatic gauge potentials that suppresses any transitions between the eigenstates back in the laboratory frame can consequently be written as
\begin{equation}
\mathcal{H}^{\lambda, \gamma}_\mathrm{CD}(t) = \dot{\lambda}(t) \mathcal{A}_\lambda(t) + \dot{\gamma}(t) \mathcal{A}_\gamma(t).
\label{App_A_eq_Hcd}
\end{equation}
It is straightforward to verify that the two adiabatic gauge potentials fulfill the relations
\begin{align}
[\mathcal{A}_{\lambda},\mathcal{H}^{\lambda, \gamma}_0] &= i\partial_{\lambda}\mathcal{H}^{\lambda, \gamma}_0 + i\mathcal{M}_{\lambda}, \nonumber \\
[\mathcal{A}_{\gamma},\mathcal{H}^{\lambda, \gamma}_0] &= i\partial_{\gamma}\mathcal{H}^{\lambda, \gamma}_0 + i\mathcal{M}_{\gamma},
\label{App_A_eq_conditions}
\end{align}
where the operators $\mathcal{M}_{\lambda}=-\sum_{n} |n\rangle \langle n| \partial_{\lambda}\mathcal{H}^{\lambda, \gamma}_0|n\rangle \langle n|$ and $\mathcal{M}_{\gamma}=-\sum_{n} |n\rangle \langle n| \partial_{\gamma}\mathcal{H}^{\lambda, \gamma}_0|n\rangle \langle n|$ are diagonal in the instantaneous eigenbasis $|n(\lambda,\gamma)\rangle$. 
As $[\mathcal{H}^{\lambda, \gamma}_0, i \mathcal{M_{\lambda}}] = [\mathcal{H}^{\lambda, \gamma}_0, i \mathcal{M_{\gamma}}] = 0$ and thus commute, we can rewrite the conditions, given by Eq.~\eqref{App_A_eq_conditions}, as
\begin{align}
[\mathcal{H}^{\lambda, \gamma}_0,\, [\mathcal{A}_{\lambda},\mathcal{H}^{\lambda, \gamma}_0] - i \partial_{\lambda}\mathcal{H}^{\lambda, \gamma}_0]  &= 0, \nonumber \\
[\mathcal{H}^{\lambda, \gamma}_0, \, [\mathcal{A}_{\gamma},\mathcal{H}^{\lambda, \gamma}_0] - i \partial_{\gamma}\mathcal{H}^{\lambda, \gamma}_0] &= 0.
\label{App_A_eq_conditions_2}
\end{align}
The exact solution for the adiabatic gauge potentials $\mathcal{A}_\lambda$ and $\mathcal{A}_\gamma$ generally requires {\em a priori} knowledge of the system eigenstates, i.e., $\mathcal{M}_\lambda$ and $\mathcal{M}_\gamma$, during the whole annealing time through $|n\rangle = |n[\lambda(t), \gamma(t)]\rangle$. To generate the latter, $\mathcal{A}_\lambda$ and $\mathcal{A}_\gamma$ have complicated many-body interacting terms of all combinations of the operators $\sigma_i^x$, $\sigma_i^y$, and $\sigma_i^z$ up to complicated nonlocal $N$-spin terms (\cf Ref.~\cite{delcampo2012assisted} in the case of quantum criticality).

\par 

To circumvent this difficulty, we follow Ref.~\cite{sels2017minimizing} and define the Hermitian operators $G_\lambda(\mathcal{A}'_\lambda) \equiv \partial_\lambda \mathcal{H}^{\lambda, \gamma}_0 + i [\mathcal{A}'_\lambda, \mathcal{H}^{\lambda, \gamma}_0]$ and $G_\gamma(\mathcal{A}'_\gamma) \equiv \partial_\gamma \mathcal{H}^{\lambda, \gamma}_0+ i [\mathcal{A}'_\gamma, \mathcal{H}^{\lambda, \gamma}_0]$ and insert a suitable \emph{Ansatz} $\mathcal{A}'_\lambda$ and $\mathcal{A}'_\gamma$, respectively, to approximately solve Eqs.~\eqref{App_A_eq_conditions_2}. Notice that inserting the exact solutions into the Hermitian operators by multiplying Eqs.~\eqref{App_A_eq_conditions} with the imaginary number $i$ and solving for the generalized forces $\mathcal{M}_\lambda$ and $\mathcal{M}_\gamma$ leads to the expressions $G_\lambda(\mathcal{A}_\lambda)=-\mathcal{M}_\lambda$ and $G_\gamma(\mathcal{A}_\gamma)=-\mathcal{M}_\gamma$.

\par 

We aim to approximate the exact solutions for the adiabatic gauge potentials as faithfully as possible. To measure the distance between our approximate ($\mathcal{A}'_\lambda$ and $\mathcal{A}'_\gamma$) and exact ($\mathcal{A}_\lambda$ and $\mathcal{A}_\gamma$) adiabatic gauge potentials, it is convenient to introduce the operator distance as the Frobenius norm.
The two-parameter operator distance can be written as
\begin{align}
&\mathcal{D}^2 = \mathrm{Tr}[ (G_{\lambda}(\mathcal{A}'_\lambda) + \mathcal{M}_{\lambda})^2] + \mathrm{Tr}[ (G_{\gamma}(\mathcal{A}'_\gamma)+ \mathcal{M}_{\gamma})^2] \nonumber \\
&= \mathrm{Tr}[G^2_\lambda(\mathcal{A}'_\lambda)] + \mathrm{Tr}[G^2_\gamma(\mathcal{A}'_\gamma)] - \mathrm{Tr}[\mathcal{M}_{\lambda}^2] - \mathrm{Tr}[\mathcal{M}_{\gamma}^2]
\label{App_A_eq_operatordistance}
\end{align}
where we use the fact that $\mathcal{H}^{\lambda, \gamma}_0$ commutes with $\mathcal{M}_\lambda$ and $\mathcal{M}_\gamma$, respectively, and  $\mathrm{Tr}[\mathcal{M}_{\lambda} \partial_{\lambda} \mathcal{H}^{\lambda, \gamma}_0]=-\mathrm{Tr}[\mathcal{M}^2_{\lambda}]$ and $\mathrm{Tr}[\mathcal{M}_{\gamma} \partial_{\gamma} \mathcal{H}^{\lambda, \gamma}_0]=-\mathrm{Tr}[\mathcal{M}^2_{\gamma}]$. As the generalized forces $\mathcal{M}_\lambda$ and $\mathcal{M}_\gamma$ do not depend on $\mathcal{A}'_\lambda$ and $\mathcal{A}'_\gamma$, we can minimize the two-parameter operator distance, given by Eq.~\eqref{App_A_eq_operatordistance}, by minimizing the two-parameter action
\begin{equation}
\mathcal{S}=\mathrm{Tr}[G^2_\lambda(\mathcal{A}'_\lambda)] + \mathrm{Tr}[G^2_\gamma(\mathcal{A}'_\gamma)]
\label{App_A_eq_action}
\end{equation}
with respect to the parameters of our \emph{Ans\"atze} for the adiabatic gauge potentials, $\mathcal{A}'_\lambda$ and $\mathcal{A}'_\gamma$, symbolically written as $\{\delta \mathcal{S}/\delta \mathcal{A}'_\lambda=0, \, \delta \mathcal{S}/\delta \mathcal{A}'_\gamma=0\}$.

\section{Landau-Zener model}\label{App_B}
In this appendix, we illustrate the method of our two-parameter CD drive for the Landau-Zener model. Its original Hamiltonian reads
\begin{equation}
\mathcal{H}^{\lambda, \gamma}_\mathrm{LZ,0}(t)=-[1-\lambda(t)] \gamma(t) \sigma^x - \lambda(t) h \sigma^z,
\label{App_B_eq_H0_LZ}
\end{equation}
where the driving functions are
\begin{align}
\lambda(t) &= \sin^2\left[\dfrac{\pi}{2}\sin^2\left(\dfrac{\pi t}{2 \tau}\right)\right], \nonumber \\
\gamma(t) &= \gamma_\mathrm{init} + \lambda(t).
\label{App_B_eq_driving_functions}
\end{align}
We have followed Ref.~\cite{sels2017minimizing} in choosing the functional form of $\lambda (t)$ and have arbitrarily chosen the form of $\gamma(t)$.
We now employ the \emph{Ans\"atze} $\mathcal{A}'_\lambda \equiv \alpha \sigma^y$ and $\mathcal{A}'_\gamma \equiv \beta \sigma^y$ for the adiabatic gauge potentials with respect to $\lambda$ and $\gamma$, respectively, and calculate the two Hermitian operators $G_\lambda(\mathcal{A}'_\lambda)=\partial_\lambda \mathcal{H}^{\lambda, \gamma}_\mathrm{LZ,0} + i [\mathcal{A}'_\lambda, \mathcal{H}^{\lambda, \gamma}_\mathrm{LZ,0}]$ and $G_\gamma(\mathcal{A}'_\gamma)=\partial_\gamma \mathcal{H}^{\lambda, \gamma}_\mathrm{LZ,0} + i [\mathcal{A}'_\gamma, \mathcal{H}^{\lambda, \gamma}_\mathrm{LZ,0}]$ and then minimize the corresponding two-parameter action $\mathcal{S}$, given by Eq.~\eqref{App_A_eq_action}, with respect to the coefficients $\alpha$ and $\beta$.
The Hermitian operators then turn out to be
\begin{align}
G_\lambda(\mathcal{A}'_\lambda)&=(\gamma + 2\lambda h \alpha) \sigma^x - [h + 2(1-\lambda)\gamma\alpha] \sigma^z, \nonumber \\
G_\gamma(\mathcal{A}'_\gamma)&=[2\lambda h \beta - (1-\lambda)]\sigma^x - 2(1-\lambda)\gamma\beta\sigma^z.
\label{eq_App_B_Hermitianoperators_LZ}
\end{align}
Pauli matrices are traceless and thus calculating the trace of the square of the Hermitian operators is equivalent to adding up squares of the coefficients in front of every Pauli matrix.
Therefore, the action $\mathcal{S}$, given by Eq.~\eqref{App_A_eq_action}, reads
\begin{align}
\mathcal{S}&=(\gamma + 2\lambda h \alpha)^2 + [h + 2(1-\lambda)\gamma\alpha]^2 \nonumber \\
&+[2\lambda h \beta - (1-\lambda)]^2 + 4(1-\lambda)^2 \gamma^2 \beta^2.
\label{eq_App_B_action_LZ}
\end{align}
By calculating the derivatives of this action with respect to $\alpha$ and $\beta$, i.e., solving the system of equations $\{\delta \mathcal{S}/\delta \alpha=0, \, \delta \mathcal{S}/\delta \beta=0 \}$, we obtain the optimal solution for the CD Hamiltonian $\mathcal{H}^{\lambda, \gamma}_\mathrm{LZ,CD}(t)=(\dot{\lambda} \alpha + \dot{\gamma} \beta )\sigma^y$, given by Eq.~\eqref{eq_H_CD_new} from the main text, as
\begin{align}
\alpha &= -\dfrac{1}{2} \dfrac{h \gamma(t)}{\lambda^2(t) h^2 + \gamma^2(t) (1-\lambda(t))^2}, \nonumber \\
\beta &= \dfrac{1}{2} \dfrac{[1-\lambda(t)] \lambda(t) h}{\lambda^2(t) h^2 + \gamma^2(t) (1-\lambda(t))^2}.
\label{eq_solution_LZ}
\end{align}
It turns out that this solution reduces to the exact CD term (\cf Ref.~\cite{takahashi2013transitionless}) when $\gamma(t)$ is constant--- as $\dot{\gamma}(t)$ then becomes zero and, consequently, we are left with the solution for $\alpha$ in Eq.~\eqref{eq_solution_LZ} with $\gamma(t) = \gamma$ alone, i.e., $\beta = 0$.

\par 

We can gauge away the imaginary $\sigma^y$ term by applying the unitary gauge transformation $U_\mathrm{g}(t)=\exp[i \theta(t) \sigma^z / 2] = \cos(\theta(t)/2) \mathds{1} + i \sin(\theta(t)/2) \sigma^z$ to the full Hamiltonian $\mathcal{H}^{\lambda, \gamma}_\mathrm{LZ}(t) = \mathcal{H}^{\lambda, \gamma}_\mathrm{LZ,0}(t) + \mathcal{H}^{\lambda, \gamma}_\mathrm{LZ,CD}(t)$ according to
\begin{equation}
\mathcal{H}^{\lambda, \gamma}_\textrm{LZ,eff}(t)=U_\mathrm{g} \mathcal{H}^{\lambda, \gamma}_\textrm{LZ}(t) U^{\dagger}_\mathrm{g} + i (\partial_t U_\mathrm{g}) U^{\dagger}_\mathrm{g}. 
\label{App_B_eq:rotated_Hamiltonian}
\end{equation} 
Here, the second term evaluates to $i (\partial_t U_\mathrm{g}) U^{\dagger}_\mathrm{g}=-(\dot{\theta}/2) \sigma^z$ with the right angle $\theta=\arctan(Y/X)$ along with $X=-[1 -\lambda(t)] \gamma(t)$ and $Y= \dot{\lambda}(t) \alpha(t) + \dot{\gamma}(t) \beta(t)$ and where we set $\hbar=1$. 
The effective full Hamiltonian in the rotated frame then reads
\begin{equation}
    \mathcal{H}^{\lambda, \gamma}_\mathrm{LZ,eff}(t) = \sqrt{X^2 + Y^2} \sigma^x - \left[\dfrac{1}{2} \dfrac{\dot{Y} X - \dot{X} Y}{X^2 + Y^2} + h \lambda(t) \right] \sigma^z,
    \label{AppB_eq_H_LZ_full}
\end{equation}
where the term involving time derivatives of $X$ and $Y$ stems from the corresponding derivatives of $U_\mathrm{g}$ and $\dot{\theta}$.

\par 

\begin{figure}
  \centering
  \includegraphics[width=.95\columnwidth]{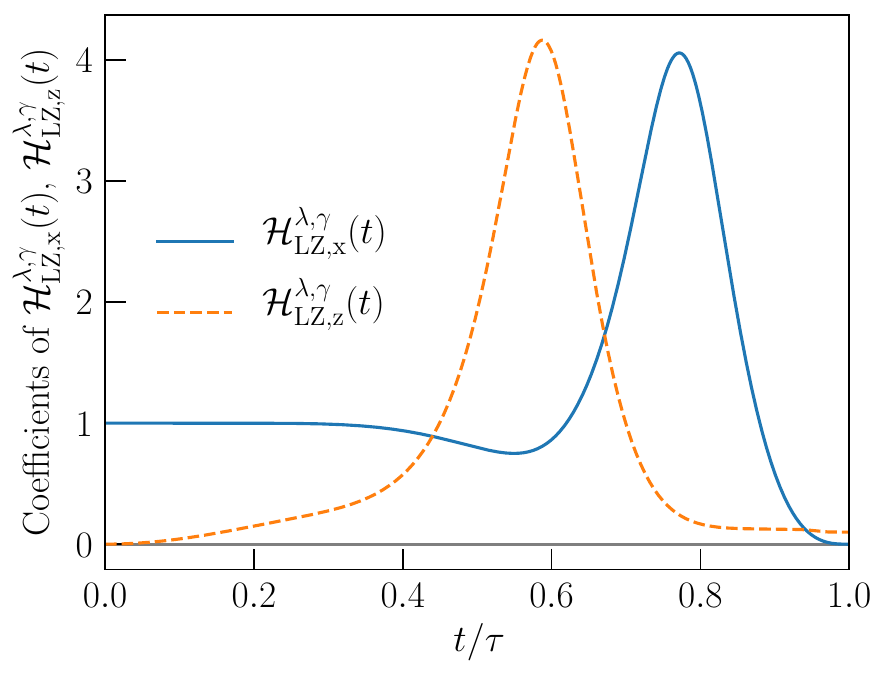}
  \caption{\textbf{Coefficient of two-parameter CD drive.} Coefficients of the driver Hamiltonian, i.e., $\mathcal{H}^{\lambda,\gamma}_\mathrm{LZ,x}(t)$ (blue solid line) and problem Hamiltonian $\mathcal{H}^{\lambda,\gamma}_\mathrm{LZ,z}(t)$ (orange dashed line) as described in Eq.~\eqref{App_B_eq_rotatedframe} for two-parameter CD driving during annealing of $\tau=1$. Other parameter: $h=0.1$.
  }
  \label{fig:coefficients_LZ}
\end{figure}

Figure~\ref{fig:coefficients_LZ} depicts the coefficients of the rotated driver and problem Hamiltonian, i.e.,
\begin{align}
    \mathcal{H}^{\lambda,\gamma}_\mathrm{LZ,x}(t)&=\sqrt{X^2 + Y^2} \sigma^x, \nonumber \\
    \mathcal{H}^{\lambda,\gamma}_\mathrm{LZ,z}(t)&=- \left[\dfrac{ \dot{Y} X- \dot{X} Y}{2(X^2 + Y^2)} + h \lambda(t)\right] \sigma^z,
    \label{App_B_eq_rotatedframe}
\end{align}
with $\gamma(t)=\gamma_\mathrm{init} + \lambda(t)$ and annealing time $\tau=1$. The coefficients are quite non-monotonic and become rather large at intermediate times.

Finally, we note that an introduction of $\sigma^x$ and $\sigma^z$ in the approximate adiabatic gauge potential, in addition to $\sigma^y$, i.e., employing the \emph{Ans\"atze} $\mathcal{A}'_\lambda = \alpha^x \sigma^x + \alpha^y \sigma^y + \alpha^z \sigma^z$ and $\mathcal{A}'_\gamma = \beta^x \sigma^x + \beta^y \sigma^y + \beta^z \sigma^z$, leads to vanishing coefficients $\alpha^x$ and $\alpha^z$ as well as $\beta^x$ and $\beta^z$. This can directly be seen by calculating the corresponding Hermitian operators $G(\mathcal{A}_\lambda)$ and $G(\mathcal{A}_\gamma)$, which entail additional $2[(1-\lambda)\gamma \alpha^z - \lambda h \alpha^x]\sigma^y$ and $2[(1-\lambda)\gamma \beta^z - \lambda h \beta^x]\sigma^y$ terms. The resulting action $S$, given by Eq.~\eqref{App_A_eq_action}, thus comprises additional $(\alpha^x)^2$, $(\alpha^z)^2$, and $\alpha^x \alpha^z$ as well as $(\beta^x)^2$, $(\beta^z)^2$, and $\beta^x \beta^z$ terms which, after taking the square and building the derivative with respect to $\alpha^x$ and $\alpha^z$ as well as $\beta^x$ and $\beta^z$, become zero.
This justifies the framework to use only $\sigma^y$ in the approximate adiabatic gauge potential. The same can be observed in more generic cases with interactions in the cost function, i.e., the Ising model.

\section{\pmb{$p$}-spin model}\label{App_C}
In this appendix, we derive the solutions of the optimal two-parameter CD Hamiltonian, $\mathcal{H}^{\lambda, \gamma}_\mathrm{CD}(t)=\sum_{i=1}^N (\dot{\lambda} \alpha + \dot{\gamma} \beta) \sigma_i^y$, given by Eq.~\eqref{eq_solution_p_spin} from the main text, for the $p$-spin model with $p=3$ and original Hamiltonian $\mathcal{H}_0(t)$, given by Eq.~\eqref{eq_H0_p_spin}, with driving functions $\lambda(t)$ and $\gamma(t)$, given by Eq.~\eqref{eq_driving_functions}. 
For the latter, we can rewrite the original Hamiltonian into the form
\begin{align}
\mathcal{H}^{\lambda, \gamma}_0(t) &= -[1 - \lambda(t)] \sum_{i=1}^N \gamma(t) \sigma_i^x  \nonumber \\
&- \lambda(t) \frac{1}{N^2} \left[6 \sum_{i< j< k} \sigma^z_i \sigma^z_j \sigma^z_k + (3N-2) \sum^N_{i=1} \sigma^z_i \right].
\label{App_C_eq_H0_p_spin}
\end{align}
For this many-body case, we employ the \emph{Ans\"atze} $\mathcal{A}'_\lambda \equiv \sum_{i=1}^N \alpha \sigma_i^y$ and $\mathcal{A}'_\gamma \equiv \sum_{i=1}^N \beta \sigma_i^y$ for the corresponding adiabatic gauge potentials. Calculating the Hermitian operators $G_\lambda(\mathcal{A}'_\lambda)$ and $G_\gamma(\mathcal{A}'_\gamma)$ requires the commutators
\begin{align}
i [\mathcal{A}'_\lambda, \mathcal{H}^{\lambda, \gamma}_0]&= \sum_{i=1}^N \frac{2\lambda(3N-2)}{N^2}\alpha\sigma^x_i - 2(1-\lambda)  \alpha \gamma \sigma^z_i \nonumber \\ 
&+ \frac{12\lambda}{N^2}\sum^N_{i< j< k} \alpha (\sigma_i^x \sigma_j^z  \sigma_k^z + \sigma_i^z \sigma^x_j \sigma^z_k + \sigma_i^z \sigma_j^z \sigma_k^x), \nonumber \\
i [\mathcal{A}'_\gamma, \mathcal{H}^{\lambda, \gamma}_0]&= \sum_{i=1}^N \frac{2\lambda(3N-2)}{N^2}\beta\sigma^x_i - 2(1-\lambda)  \beta \gamma \sigma^z_i \nonumber \\ 
&+ \frac{12\lambda}{N^2}\sum^N_{i< j< k} \beta (\sigma_i^x \sigma_j^z  \sigma_k^z + \sigma_i^z \sigma^x_j \sigma^z_k + \sigma_i^z \sigma_j^z \sigma_k^x).
\end{align}
Adding the two partial derivatives $\partial_\lambda \mathcal{H}^{\lambda, \gamma}_0$ and $\partial_\gamma \mathcal{H}^{\lambda, \gamma}_0$, respectively, leads to the Hermitian operators
\begin{align}
G_\lambda(\mathcal{A}'_\lambda)&=\sum_{i=1}^N \left[\gamma + \frac{2\alpha\lambda(3N-2)}{N^2}\right] \sigma_i^x - \frac{6}{N^2} \sum^N_{i< j< k} \sigma^z_i \sigma^z_j \sigma^z_k \nonumber \\ 
&- \sum^N_{i=1} \left[\frac{3N-2}{N^2} + 2\alpha(1-\lambda)\gamma \right] \sigma^z_i \nonumber \\
&+ \frac{12\lambda}{N^2}\sum^N_{i< j< k} \alpha (\sigma_i^x \sigma_j^z  \sigma_k^z + \sigma_i^z \sigma^x_j \sigma^z_k + \sigma_i^z \sigma_j^z \sigma_k^x), \nonumber \\
G_\gamma(\mathcal{A}'_\gamma)&=\sum_{i=1}^N \left[\frac{2\beta\lambda(3N-2)}{N^2} - (1 - \lambda) \right] \sigma_i^x \nonumber \\ 
&- \sum^N_{i=1} \left[\frac{3N-2}{N^2} + 2\beta(1-\lambda)\gamma \right] \sigma^z_i - \frac{6}{N^2} \sum^N_{i< j< k} \sigma^z_i \sigma^z_j \sigma^z_k \nonumber \\
&+ \frac{12\lambda}{N^2}\sum^N_{i< j< k} \beta (\sigma_i^x \sigma_j^z  \sigma_k^z + \sigma_i^z \sigma^x_j \sigma^z_k + \sigma_i^z \sigma_j^z \sigma_k^x).
\label{eq_App_C_Hermitianoperators_p_spin}
\end{align}
Consequently, the action $\mathcal{S}=\mathrm{Tr}[G^2_\lambda(\mathcal{A}'_\lambda)] + \mathrm{Tr}[G^2_\gamma(\mathcal{A}'_\gamma)]$, given by Eq.~\eqref{App_A_eq_action}, can be written as
\begin{widetext}
\begin{align}
\dfrac{\mathcal{S}}{2^N}&= \sum_{i=1}^N \left[\gamma + \frac{2\alpha\lambda(3N-2)}{N^2}\right]^2 + \left[ \frac{3N-2}{N^2} + 2\alpha(1-\lambda)\gamma \right]^2 +  \frac{72\lambda^2}{N^4} (N-1)(N-2) \alpha^2 \nonumber \\
&+\sum_{i=1}^N \left[\frac{2\beta\lambda(3N-2)}{N^2} - (1 - \lambda) \right]^2 + \left[ \frac{3N-2}{N^2} + 2\beta(1-\lambda)\gamma \right]^2 +  \frac{72\lambda^2}{N^4} (N-1)(N-2) \beta^2 - \frac{12(N-1)(N-2)}{N^2} 
\label{App_C_eq_action_p_spin}
\end{align}
\end{widetext}
and minimizing this action with respect to each coefficient $\alpha$ and $\beta$ leads to the solutions, given by Eq.~\eqref{eq_solution_p_spin}, from the text.

\par 

To bring the full Hamiltonian $\mathcal{H}^{\lambda, \gamma}(t) = \mathcal{H}^{\lambda, \gamma}_0(t) + \mathcal{H}^{\lambda, \gamma}_\mathrm{CD}(t)$ with $\mathcal{H}^{\lambda, \gamma}_\mathrm{CD}(t)$ from Eq.~\eqref{eq_solution_p_spin} in an experimentally more feasible form, we can gauge away the imaginary single-body $\sigma_i^y$ terms by applying the unitary gauge transformation $U_\mathrm{g}[\theta(t)] = \exp[i \theta(t) /2 \sum_{i=1}^N \sigma_i^z]$, given by Eq.~\eqref{eq:unitary_transformation}, for convenience written as $U_\mathrm{g} = \prod_{j=1}^N [\cos(\theta/2) \mathds{1} + i \sin(\theta/2) \sigma_j^z]$. 
The effective, i.e., rotated, full Hamiltonian in the laboratory frame then reads
\begin{equation}
\mathcal{H}^{\lambda, \gamma}_\textrm{eff}(t)=U_\mathrm{g} \mathcal{H}^{\lambda, \gamma}(t) U^{\dagger}_\mathrm{g} - \sum_{i=1}^N \dfrac{\dot{\theta}}{2} \sigma_i^z
\label{App_C_eq:rotated_Hamiltonian}
\end{equation}
which can straightforwardly be derived by multiplying both sides of the time-dependent Schr\"odinger equation $i \partial_t \psi = H \psi$ in the original frame with the unitary transformation $U_\mathrm{g}$, given by Eq.~\eqref{eq:unitary_transformation}, employing the relation $\tilde{\psi} = U_\mathrm{g} \psi$ and expressing the dynamics in the moving frame.
Analogously to Appendix~\ref{App_B}, the rotational right angle is $\theta = \arctan(Y/X)$ with $Y = \dot{\lambda} \alpha + \dot{\gamma} \beta$ and $X = - (1 - \lambda) \gamma$. 
For the first term of Eq.~\eqref{App_C_eq:rotated_Hamiltonian}, we use that $U_\mathrm{g} \sigma_i^x U^\dagger_\mathrm{g} = \cos \theta \sigma_i^x - \sin \theta \sigma_i^y$, $U_\mathrm{g} \sigma_i^y U^\dagger_\mathrm{g} = \sin \theta \sigma_i^x + \cos \theta \sigma_i^y$, and $U_\mathrm{g} \sigma_i^z U^\dagger_\mathrm{g} = \sigma_i^z$, as well as the trigonometrical relations $\sin \theta = Y / \sqrt{X^2 + Y^2}$ and $\cos \theta = X / \sqrt{X^2 + Y^2}$. The first term then evaluates to $\sqrt{X^2 + Y^2} \sigma_i^x$ and together with the second term and $\dot{\theta} = (\dot{Y} X - \dot{X} Y)/(X^2 + Y^2)$ describe the expression of the effective Hamiltonian, given by Eq.~\eqref{eq_H_full_eff}, from the text.

\section{Different driving functions $\pmb{\gamma_i(t)}$}
\label{appendix:different_gamma}
\begin{figure*}[htbp]
  \includegraphics[width=.99\textwidth]{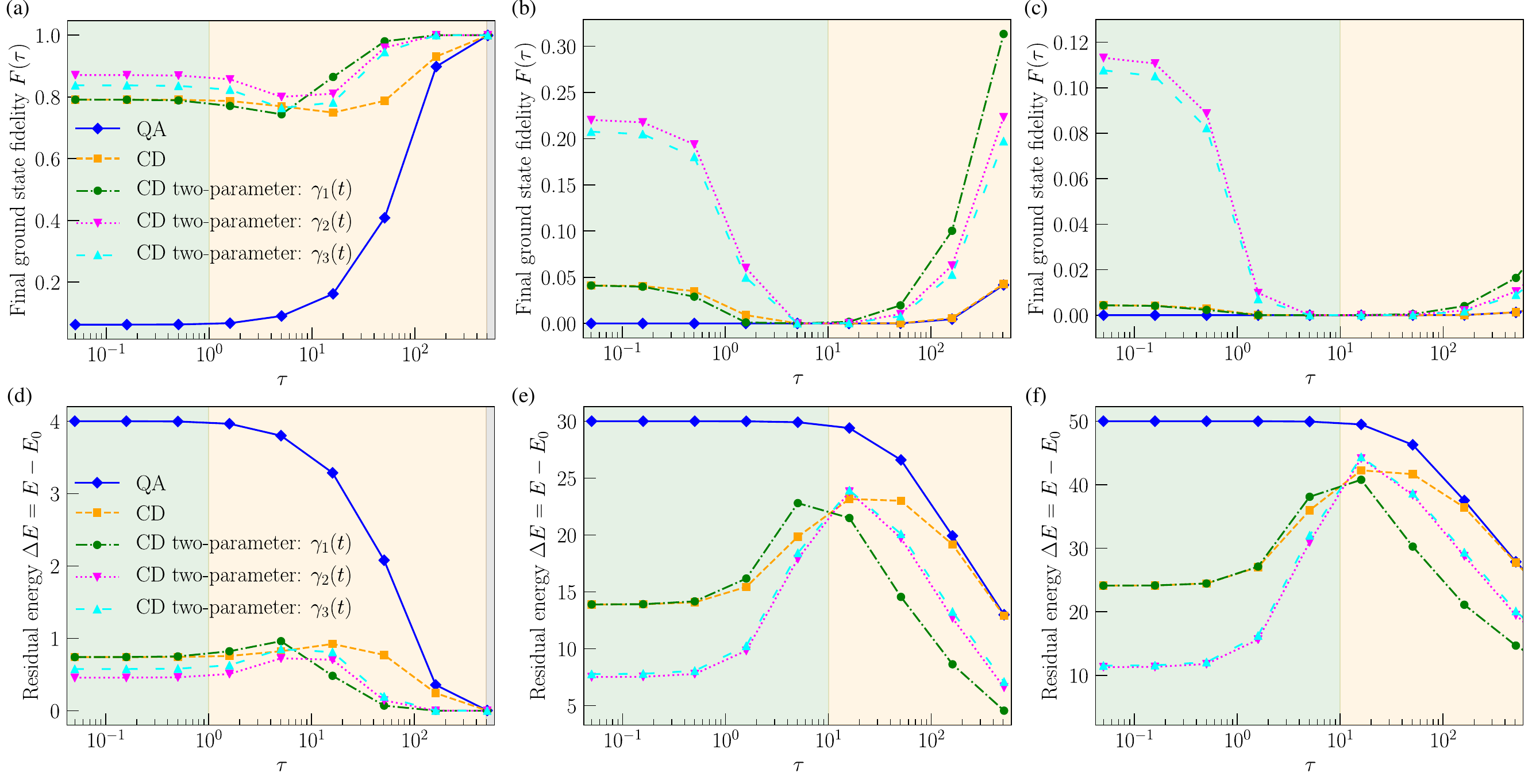}
  \caption{\textbf{Ground-state fidelity and residual energy.} (a)--(c)~Final ground-state fidelity and (d)-(f)~residual energy for (i)~traditional quantum annealing (diamonds, blue solid line), (ii)~single-parameter CD drive (squares, orange dashed line), and (iii)~two-parameter CD drive with driving functions $\gamma_1(t)$, given by Eq.~\eqref{eq:gamma1} (circles, green dash-dotted line), $\gamma_2(t)$, given by Eq.~\eqref{eq:gamma2} (down triangles, magenta dotted line), and $\gamma_3(t)$, given by Eq.~\eqref{eq:gamma3} (up triangles, cyan widely dashed line), as functions of annealing time $\tau$. The system sizes are  (a),(d) $N=4$, (b),(e) $N=30$, and (c),(f) $N=50$, where $\gamma_\mathrm{init}=1$ for all panels. Time ranges are color coded as follows: short-time regime (green-shaded areas) where the fidelity is approximately $1/2^N$ for traditional quantum annealing, long-time regime (yellow-shaded areas) where transient behavior is observed and the two-parameter CD drive shows a clear advantage, and adiabatic regime (gray-shaded areas) where $F(\tau) > 0.99$.}
  \label{fig:Appendix_D}
\end{figure*}

The numerical results of the two-parameter CD approach with particular choice of the driving function $\gamma(t)$, given by Eq.~\eqref{eq_driving_functions}, revealed a considerable enhancement in the reached final ground-state fidelity and residual energy.
We are thus interested in whether this enhancement stems from this particular choice of driving functions, or constitutes a general feature due to the expansion of the search space for the optimal parameters $\alpha$ and $\beta$, given by Eq.~\eqref{eq_solution_p_spin}. 
Although it is difficult to systematically explore the best possible functional forms, we nevertheless tried a few different cases to confirm that our conclusion remains unchanged qualitatively.
To this end, we compare the numerical performance of this two-parameter CD method for three different forms of the driving function, i.e. 
\begin{align}
    \lambda(t)&=\sin^3\left( \dfrac{\pi t}{2 \tau} \right),\\
    \gamma_1(t) &= \gamma_{\mathrm{init}} - \lambda(t), \label{eq:gamma1}\\
    \gamma_2(t) &= \cos^3 \left( \dfrac{\pi t}{2 \tau} \right),\label{eq:gamma2} \\
    \gamma_3(t) &= 1 - \sin^3 \left( \dfrac{\pi t}{2 \tau} \right)\label{eq:gamma3}
\end{align}
where, in contrast to the case of Fig.~\ref{fig:p_spin} we set the initial value of $\gamma(t)$ to $\gamma_{\mathrm{init}} = 1$.
The corresponding numerical results are depicted in Fig.~\ref{fig:Appendix_D} with the same other parameters as in Fig.~\ref{fig:p_spin}. They reveal that the full Hamiltonians with two-parameter CD driving and all three driving functions considerably outperform the traditional quantum annealing and existing one-parameter counterparts. Interestingly, the two newly added driving functions $\gamma_2(t)$ and $\gamma_3(t)$, which have considerably different forms than the one originally used in Fig.~\ref{fig:p_spin}, even considerably outperform the latter for short sweep durations (green-shaded area), though the function $\gamma_1(t)$, which is similar to the one in the main text, works best in the intermediate time region (yellow-shaded area).
This is a promising result as the two-parameter approach provides a systematic enhancement for a variety of driving functions due to the expanded search space in two dimensions. These results motivate more systematic analytical and numerical investigations of driving functions that yield the maximal reached final ground-state fidelities and--- as mentioned in the main text--- constitute an interesting topic for future research.

\end{document}